# Thermal noise in half infinite mirrors with non-uniform loss: a slab of excess loss in a half infinite mirror

## N. Nakagawa

Center for Nondestructive Evaluation, Iowa State University, Ames, Iowa 50011

## A. M. Gretarsson

Physics Department, Syracuse University, Syracuse, New York 13244-1130

## E.K. Gustafson and M.M. Fejer

Ginzton Laboratory, Stanford University, Stanford, California 94305-4085

We calculate the thermal noise in half-infinite mirrors containing a layer of arbitrary thickness and depth made of excessively lossy material but with the same elastic material properties as the substrate. For the special case of a thin lossy layer on the surface of the mirror, the excess noise scales as the ratio of the coating loss to the substrate loss and as the ratio of the coating thickness to the laser beam spot size. Assuming a silica substrate with a loss function of $3\times10^{-8}$ the coating loss must be less than $3\times10^{-5}$ for a 6 cm spot size and a 7 μm thick coating to avoid increasing the spectral density of displacement noise by more than 10%. A similar number is obtained for sapphire test masses.

## Introduction

The second generation LIGO gravitational wave detector will require large core optics with low internal dissipation. The baseline design for LIGO II will use 30 kg sapphire mirrors; a fall back design would use silica mirrors. (1) A loss function less than $3.3\times10^{-9}$ has been measured at high frequencies in small samples of sapphire (2, 3). Substrate loss functions for silica have been measured to be around or below $3\times10^{-8}$ in several different sample geometries (4,5,6,7). However, to achieve this low loss in a full size coated and suspended LIGO II mirror the attachments for the fused silica suspension and the multilayer mirror high reflector coating and the antireflection optical coating must not increase the displacement noise of the mirror in the frequency range of interest.

Until recently the approach used to calculate mirror thermal noise involved a normal mode expansion of the mirror acoustic modes (8). Direct approaches to the problem have been developed by Levin (9), Nakagawa et al. (10,11), Bondu et al. (12) and Liu and Thorne (13). Moreover, Levin has used this approach to point out that non-uniform loss in a mirror can lead to higher than otherwise expected thermal noise.(9) In this paper we use the general formalism developed in (10) and (11) to derive expressions for the phase noise imposed on a Gaussian light beam when it is reflected from a half-infinite lossless mirror with a lossy layer of arbitrary thickness placed at an arbitrary depth from the surface from which the light beam is reflected. To compute the noise for the case where the mirror has both loss and an extra layer with a different loss we take the incoherent sum of the noises from the uniform lossy mirror and the lossless mirror with a lossy layer. This procedure is legitimate for low-loss cases where the loss can be estimated accurately by up to linear terms in the loss functions. Finally, we discuss the experimentally important special case where the lossy layer is at the surface of the mirror and is thin compared to the spot size of the light beam.

## Mathematical Formalism and Review

In our previous paper (11) we used the fluctuation dissipation theorem to compute the cross spectral density of the mirror surface displacements resulting from the off-resonance thermal noise in the test masses. The loss was parameterized through the imaginary part of the stiffness tensor **c**; the calculation was carried out in the quasi-static limit for isotropic and uniform loss in which case the fluctuation dissipation theorem can be expressed in terms of the static Green's function. As we showed in (11) the power spectral density of the phase noise imposed on a light beam of a field-amplitude radius $w$ reflected from a lossy half-infinite mirror can be written

$$S_{\varphi}^{Single}(f)=\frac{16 k_B T}{\pi^3}\frac{1}{f}\frac{k^2}{w^4}\iint dS'\iint dS'' e^{-2|\vec{r}'|^2/w^2} e^{-2|\vec{r}''|^2/w^2}\,\mathrm{Im}\ \chi_{zz}^{w}(\vec{r}',\vec{r}'') \tag{1}$$

This result relates the spectral density of phase noise induced on a Gaussian beam by reflection from a mirror at temperature, $T$, to the static elastic properties of the mirror, the imaginary part of the elastic Green's function, and the laser beam spot size on the mirror.

It can be shown further that, under the usual single-loss-function assumption, the imaginary part of the elastic Green's function is proportional to the loss function and the static Green's function (cf. Appendix A). The essential input for these calculations is, therefore, the static Green's function for an isotropic half-space medium given in Ref. (14),

$$\mathrm{Im}\ \chi_{zz}^{w}(\vec{r}',\vec{r}'')=\phi\cdot\chi_{zz}^{st}(\vec{r}',\vec{r}'')=\phi\frac{1-\sigma^2}{\pi E}\frac{1}{|\vec{r}'-\vec{r}''|} \tag{2}$$

where $\phi$ is a loss function, $\sigma$ the Poisson ratio, $E$ the Young's modulus, and $\vec{r}'$ and $\vec{r}''$ are surface points. The phase noise after a single bounce from a half-infinite mirror with uniform loss is obtained by substituting Eq. (2) for the imaginary part of the Green's function into Eq. (1) for the phase noise

$$S_{\varphi}^{Single}(f,\phi)=\frac{4k_B T}{\pi^{3/2}}\frac{\phi(f)}{f}\frac{k^2}{w}\frac{(1-\sigma^2)}{E} \tag{3}$$

which is in agreement with Levin's result (11) for the displacement noise as corrected by Liu (13) and noting that our expression is for the double sided spectral density and Levin's result is for a single sided spectral density. We will now investigate the case where a lossy slab is embedded in a lossless half infinite mirror, for which we need the Green's function for this more complicated problem.

## Lossless half infinite mirror containing a lossy layer

To calculate the phase noise after a single bounce from a lossless mirror containing a slab with loss we start as before with the elastic Green's function. For a half-infinite mirror whose surface is at $z=0$ and all of whose loss is confined to a layer between $z_1$ and $z_2$, the imaginary part of the Green's function is derived in Appendix A,

$$\mathrm{Im}\ \chi_{zz}^{w}(\vec{r}\,' - \vec{r}\,'') = \phi \frac{1-\sigma^{2}}{\pi E}\left[F(z_1) - F(z_2)\right] \tag{4}$$

where the function $F$ is defined as

$$F(z) = \frac{1}{\sqrt{\left|\vec{r}\,' - \vec{r}\,''\right|^{2} + 4z^{2}}}\left(1 + \frac{z^{2}/(1-\sigma)}{\left(\left|\vec{r}\,' - \vec{r}\,''\right|^{2} + 4z^{2}\right)} + \frac{12\,z^{4}/(1-\sigma)}{\left(\left|\vec{r}\,' - \vec{r}\,''\right|^{2} + 4z^{2}\right)^{2}}\right) \tag{5}$$

Substituting equation (4) for the imaginary part of the Green's function into equation (2) for the phase noise after a single bounce we find

$$S_j^{Layer}(f) = \left(\frac{16 k_B T}{\pi^{3}} \frac{\phi_{layer}(f)}{f} \frac{k^{2}}{w^{4}}\right)\frac{\left(1-\sigma^{2}\right)}{\pi E}\iint dS' \iint dS' e^{-2\left(\frac{\vec{r}\,'^{2}}{w^{2}}\right)} e^{-2\left(\frac{\vec{r}\,''^{2}}{w^{2}}\right)}\left[F(z_1) - F(z_2)\right] \tag{6}$$

The two surface integrals can be performed as shown in Appendix B; inserting these results into Eq. (6) we obtain

$$\begin{aligned} S_j^{Layer}(f, w) &= \frac{4KT}{\pi^{3/2}} \frac{\phi_{Layer}(f)}{f} \frac{k^{2}}{w} \frac{\left(1-\sigma^{2}\right)}{E}\left[G(w,\sigma,z_1) - G(w,\sigma,z_2)\right] \\ &= S_j^{Single}\left(f, \phi_{layer}\right)\left[G(w,\sigma,z_1) - G(w,\sigma,z_2)\right] \end{aligned} \tag{7}$$

where in the second equation we use equation 3 for the thermal noise from a uniform half infinite mirror and where G is defined as

$$G(w,\sigma,z) = \frac{2}{\sqrt{\pi}}\int_0^{\infty} dx\, e^{-x^{2}}\left\{e^{-\frac{4z}{w}x}\left[1 + \frac{2z}{w(1-\sigma)}x + \frac{4z^{2}}{w^{2}(1-\sigma)}x^{2}\right]\right\} \quad . \tag{8}$$

This integral is evaluated analytically in Appendix B,

$$G(w,\sigma,z) = \frac{1}{\sqrt{\pi}}\frac{2}{1-\sigma}\left[(z/w) - 4(z/w)^{3}\right] + \left[1 - \frac{2(z/w)^{2}}{1-\sigma} + \frac{16(z/w)^{4}}{1-\sigma}\right] e^{4(z/w)^{2}}\,\mathrm{erfc}\,(2z/w) \quad . \tag{9}$$

The factor preceding the square brackets in Eq. (7) is the phase noise after a single bounce from a half infinite mirror with uniform loss $\phi_{layer}$. If we set $z_1 = 0$ and take the limit as $z_2$ becomes large, $G$ approaches 1 and we re-derive Eq. (3) for a mirror with uniform loss. Figure 1 plots $G(w,\sigma,z_2) - G(w,\sigma,z_1 = 0)$ as a function of $z_2/w$. We see that $G(w,\sigma,z_2) - G(w,\sigma,z_1 = 0)$ grows monotonically with the layer thickness, and approaches 1 for $z \geq w$. If we fix the layer thickness at $z_2 - z_1 = d$ and then plot the phase noise as a function of the depth of the layer below the surface we find that the phase noise goes to zero as the layer is lowered into the mirror beyond a characteristic depth approximately equal to the light beam spot size. The result of evaluating Eq. (7) for several values of $d = z_2 - z_1$ as a function of $z_1/w$ is shown in Figure 2.

### Lossless half infinite mirror containing a thin lossy surface layer

For the important practical case, which approximates a multilayer dielectric coating, as a thin lossy layer at the surface of the mirror, i.e. $z_1 = 0$, $z_2 = d$ where $d << w$, $G$ can be integrated analytically as shown in Appendix B to yield

$$S_j^{Layer}(f) = S_j^{Single}\left(f, \phi_{Layer}\right)\left\{\frac{2}{\sqrt{\pi}}\frac{(1-2\sigma)}{(1-\sigma)}\left(\frac{d}{w}\right) + O\left(\frac{d}{w}\right)^2\right\} . \qquad (10)$$

### Lossy Half infinite mirrors containing a thick lossy slab

To evaluate the case of a half-infinite lossy mirror in which is embedded a lossy layer, we use linear superposition. As shown in Appendix A, the uniform-loss contribution can be given by Eq. (3) with $\phi = \phi_{substrate}$, while the excess layer loss is given by Eq. (10) with $\phi_{layer} \to \phi_{layer} - \phi_{substrate}$ . Then by linearly superposing these two spectral densities we obtain the total phase noise imposed on the light beam.

$$S_j^{Total}(f) = S_j^{Single}\left(f, \phi_{Substrate}\right)\left\{1 - \left(1 - \frac{\phi_{layer}(f)}{\phi_{substrate}(f)}\right)\left[G(w, \sigma, z_1) - G(w, \sigma, z_2)\right]\right\} \qquad (11)$$

For cases where the layer loss is much greater than that of the substrate this expression can be simplified to the following form

$$S_j^{Total}(f) = S_j^{Single}\left(f, \phi_{Substrate}\right)\left\{1 + \frac{\phi_{layer}(f)}{\phi_{substrate}(f)}\left[G(w, \sigma, z_1) - G(w, \sigma, z_2)\right]\right\} \qquad (12)$$

We will now examine the experimentally important case where the lossy layer is thin and at the surface on the half infinite substrate.

### Lossy Half infinite mirrors containing a thin lossy surface layer

To evaluate the case of a half-infinite lossy mirror with a thin lossy surface layer or coating we compute the limit for $z_1 = 0$ and $z_2 = d << w$ as we did above. The first case consists of a half infinite mirror with loss $\phi_{substrate}$ uniformly distributed throughout, and the second case will be a half-infinite lossless mirror containing a layer of thickness $d$ at the surface, in other words, a coating, with loss $\phi_{coating}$. With Eqs. 3 and 10 we find

$$S_j^{Total}(\omega) = S_j^{Single}\left(f, \phi_{Substrate}\right)\left\{1 + \frac{2}{\sqrt{\pi}}\frac{(1-2\sigma)}{(1-\sigma)}\frac{\phi_{coating}}{\phi_{substrate}}\left(\frac{d}{w}\right)\right\} \qquad (13)$$

where the $1/w^2$ dependence of the term corresponding to the surface layer agrees with Levin's equation 20 in reference 9.

### Comparison with Levin's Method

We have also calculated this expression using the method of Levin, and find the same result. Levin's method consists of calculating the response of the test mass to a cyclic pressure distribution applied to the test mass face, where the pressure distribution has a Gaussian profile of the same width as the beam. It can be shown (9) that for frequencies far below the lowest resonant frequency of the test mass, the (double-sided) power spectrum of excitations having the same Gaussian profile as the pressure distribution is

$$S_x(f) = \frac{2KT\,U}{\pi f F^2}\phi(f) \tag{14}$$

where $F$ is the amplitude of the Gaussian pressure distribution, $U$ is the mechanical energy of the resulting test-mass response, and $\phi(f)$ is the loss angle of the test-mass response. The mechanical energy may be shown to be (12, 13)

$$U = \frac{F^2(1-\sigma^2)}{2\sqrt{\pi}wE} \tag{15}$$

where $E$ is the Young's modulus of the mirror material and $\sigma$ is Poisson's ratio. The total loss angle $\phi(f)$ has contributions from loss in the substrate and loss in the coating. For a thin coating we can write (15)

$$\phi(f) = \phi_{substrate} + \frac{\delta U\, d}{U}\phi_{coating} \tag{16}$$

where $\delta U$ is the energy density at the surface integrated over the surface and $d$ is the thickness of the coating. Using the solutions to the elastic equations presented by Bondu (12) (as corrected by Liu (13)) we have found

$$\delta U = \frac{F^2(1+\sigma)(1-2\sigma)}{\pi w^2 E}\,. \tag{17}$$

Combining these equations, we obtain

$$S_x(f) = \frac{KT(1-\sigma^2)}{\pi^{3/2} f\, wE}\phi_{substrate}\left\{1 + \frac{2}{\sqrt{\pi}}\frac{(1-2\sigma)}{(1-\sigma)}\frac{\phi_{coating}}{\phi_{substrate}}\left(\frac{d}{w}\right)\right\} \tag{18}$$

The phase shift induced by a displacement $x$ of the mirror surface is $2kx$, where $k$ is the wavenumber, so multiplying $S_x(f)$ by $(2k)^2$ we obtain the phase noise power spectrum, $S_j^{Single}(f)$. The result is identical with Eq. 13.

## LIGO II Coating Loss Requirements for Sapphire and Silica Substrates

The frequency range over which the thermal noise in the mirror coating is important for LIGO II is between 200 and 400 Hz. In this region the displacement noise in the silica substrates is dominated by thermal noise while in sapphire the displacement noise is dominated by thermoelastic noise. At 300 Hz the sapphire linear spectral density of thermo-elastic noise is between 1/2 and 1/3 the thermal noise in silica. Considering a silica substrate the requirement for LIGO II is that neither the attachments nor the optical coating increase the power spectral density of the thermal noise by more than 10%, which puts a limit on the coating loss of

$$f_{coating} < \frac{\sqrt{p}}{20}\left(\frac{w}{d}\right)\left[\frac{(1-s)}{(1-2s)}\right]f_{sub} \qquad (19)$$

Assuming a spot size, $w = 6.0$ cm , a coating thickness $d = 7$ $m$m , and using the material properties of fused silica with a Poisson ratio of $s = 0.16$ and a loss function of $f_{sub} = 3.3 \times 10^{-8}$ we find from Eq. 19

$$f_{coating} < 3 \times 10^{-5} \qquad (20)$$

in order that the high reflectivity coating not increase the power spectral density of the thermal noise by more than 10 %. Preliminary measurements suggest that currently available coatings do not meet this requirement (5). A number in the range of $10^{-5}$ should also apply for sapphire. In both cases it should be borne in mind that we have neglected the differences in the elastic properties of the substrate and the film, an issue that will be addressed in a future paper. Preliminary calculations indicate that these corrections will not change the result by more than a few tens of percent.

## Conclusion

We have calculated the thermal noise caused by the inclusion of a lossy layer in an otherwise lossless half infinite mirror. By taking a linear superposition of this noise source with that of a uniformly lossy half infinite mirror we have computed the thermal noise in a half-infinite lossy mirror containing a layer of excessively lossy material. The limit in which this lossy layer lies at the surface of the mirror and is much thinner than the laser spot size yields a simple analytical result, from which we can set an upper limit for the acceptable loss in the LIGO II mirror coatings of $3 \times 10^{-5}$

## Acknowledgements

This research was supported by the National Science Foundation (NSF PHY 9630172, NSF PHY 9800976, NSF PHY 9602157, and NSF PHY 9630172). We thank Yu. Levin, Sheila Rowan, Gregory M. Harry, Jim Hough, Kip Thorne and Peter Saulson for helpful discussions and encouragement.

## Appendix A

In Ref. [10], we have shown from the energy-balance relation that

$$\text{Im } c^{w}_{pq}(\vec{r}_1,\vec{r}_2) = \int_V dV \left\{ \left[ \partial_i c^{st}_{jp}(\vec{x},\vec{r}_1) \right] c''_{ijlm}(\vec{x}) \left[ \partial_l c^{st}_{mq}(\vec{x},\vec{r}_2) \right] \right\} + O(w) \tag{A1}$$

where $c^{st}_{ij}(\vec{x},\vec{x}')$ is the elastic, traction-free Green's function, $\vec{r}_1$ and $\vec{r}_2$ are surface positions, and the integral is performed over the lossy region of the mirror body, namely over the volume $V$ where the loss $c''_{ijlm}(\vec{x})$ of the elastic constant is non-vanishing. Hereafter, we assume the quasi-static limit, thus ignoring the $O(w)$ terms. Let us first introduce the usual single loss function $f$, so that

$$c''_{ijlm}(\vec{x}) = f(\vec{x}) \cdot c'_{ijlm}\ , \tag{A2}$$

where we assume the same loss function for all components of the stiffness tensor, a simplifying assumption for general cases. We next separate $f$ into a uniform loss $\bar{f}$ and fluctuation $\Delta f(\vec{x})$ around it as

$$f(\vec{x}) = \bar{f} + \Delta f(\vec{x}). \tag{A3}$$

For a localized loss, we set $\bar{f} = 0$ while $\Delta f(\vec{x}) \neq 0$ in the lossy region. For a localized excess loss, we set $\bar{f}$ constant everywhere while $\Delta f(\vec{x}) \neq 0$ in the excess-loss region. Under these assumptions, the $\bar{f}$ term in (A1) can be integrated to yield

$$\text{Im } c^{w}_{pq}(\vec{r}_1,\vec{r}_2)\Big|_{uniform} = \bar{f} \cdot c^{st}_{pq}(\vec{r}_1,\vec{r}_2) \tag{A4}$$

because, when integrated by parts with respect to $\partial_i$, the integrand becomes a delta function through the field equation

$$\partial_i \left\{ c'_{ijlm} \partial_l c^{st}_{mq}(\vec{x},\vec{r}_2) \right\} = -d_{jq} d(\vec{x} - \vec{r}_2), \tag{A5}$$

while the surface integral term vanishes due to the traction-free boundary condition imposed on $c^{st}_{ij}$. For the $\Delta f(\vec{x})$ term, assume that $\Delta f(\vec{x}) = \Delta f \neq 0$ when $\vec{x}$ is in a sub-volume *V'* while vanishing outside *V'*. The volume integral of (A1) can be similarly performed by parts, except that, in this case, it only leaves the surface integral over the bounding surface *S'* of *V'*

$$\text{Im } c^{w}_{pq}(\vec{r}_1,\vec{r}_2) = \bar{f} \cdot c^{st}_{pq}(\vec{r}_1,\vec{r}_2) + \Delta f \int_{S'} dS_i \left[ c^{st}_{jp}(\vec{x},\vec{r}_1) c'_{ijlm} \P_l\, c^{st}_{mq}(\vec{x},\vec{r}_2) \right], \tag{A6}$$

because the delta function (A5) has no contribution since $\vec{r}_2$ lies outside the integration volume *V'*. Let us apply Eq. (A6) to a lossy layer, where the mirror body occupies the semi-infinite space ($-\infty < x, y < +\infty,\ z <= 0$), while the lossy region *V'* is a near-surface layer bounded by the two planes, $z = -z_1$ and $z = -z_2$ ($-z_2 \le z \le -z_1 \le 0$). Without loss of generality, we consider the local loss case, i.e. $\bar{f} = 0$ and $\Delta f = f_{layer}$, below. (For the case of a uniform and an excess layer loss, we set $\bar{f} = f_{substrate}$ and $\Delta f = f_{layer} - f_{substrate}$, and include the previous uniform-loss result, thus deriving Eq. (10).) Specifically when *p*=*q*=*z*, Equation (A6) reads

$$\text{Im } c^{w}_{zz}(\vec{r}_1,\vec{r}_2) = \Delta f \int_{S_{xy}} dS \left\{ \left[ c^{st}_{jz}(\vec{x},\vec{r}_1) c'_{zjlm} \P_l\, c^{st}_{mz}(\vec{x},\vec{r}_2) \right]_{z=-z_1} - \left[ c^{st}_{jz}(\vec{x},\vec{r}_1) c'_{zjlm} \P_l\, c^{st}_{mz}(\vec{x},\vec{r}_2) \right]_{z=-z_2} \right\} \tag{A7}$$

where the surface integral is over the *xy*-plane. Due to the translational symmetry in the *x* and *y* directions, the integral of Eq. (A7) becomes a convolution integral. Thus, by introducing

$$c^{st}_{ij}(\vec{r},z) = \frac{1}{(2p)^2} \int d^2 p\ e^{i\vec{p}\vec{r}}\ c^{st}_{ij}(\vec{p},z), \tag{A8a}$$

$$\Phi_{ijk}(\vec{r},z) \equiv c'_{ijlm}\partial_l \chi^{st}_{mk}(\vec{r},z) = \frac{1}{(2\pi)^2}\int d^2p\, e^{i\vec{p}\vec{r}}\Phi_{ijk}(\vec{p},z) \tag{A8b}$$

we can rewrite (A7) as

$$\mathrm{Im}\ \chi^{w}_{zz}(\vec{r}_1,\vec{r}_2) = \frac{\Delta\phi}{(2\pi)^2}\int d^2p\ e^{i\vec{p}(\vec{r}_1-\vec{r}_2)}\left[\chi^{st}_{jz}(\vec{p},z_1)\Phi_{zjz}(-\vec{p},z_1) - \chi^{st}_{jz}(\vec{p},z_2)\Phi_{zjz}(-\vec{p},z_2)\right]. \tag{A9}$$

The half-space Green's function (A8a) can be found in Ref. [14]. Here, we express the results in the derivative form as

$$\chi^{st}_{\alpha\beta}(\vec{r},z) = \frac{1+\sigma}{2\pi E}\left[\frac{2}{R}\delta_{\alpha\beta} - \partial_\alpha\partial_\beta\left\{2\sigma R - (1-2\sigma)z\ln(R-z)\right\}\right], \tag{A10a}$$

$$\chi^{st}_{\alpha z}(\vec{r},z) = \frac{1+\sigma}{2\pi E}\partial_\alpha\left[(1-2\sigma)\ln(R-z) - \frac{z}{R}\right], \tag{A10b}$$

$$\chi^{st}_{z\beta}(\vec{r},z) = \frac{1+\sigma}{2\pi E}\partial_\beta\left[-(1-2\sigma)\ln(R-z) - \frac{z}{R}\right], \tag{A10c}$$

$$\chi^{st}_{zz}(\vec{r},z) = \frac{1+\sigma}{2\pi E}\left[\frac{2(1-\sigma)}{R} + \frac{z^2}{R^3}\right], \tag{A10d}$$

where $\alpha,\beta = \{x,y\}$, and where $R = \sqrt{r^2+z^2}$. Explicit differentiation reproduces the result of Ref. [14] after changing $z \to -z$, while it is easy to Fourier-transform the expressions (A10) to

$$\chi^{st}_{\alpha\beta}(\vec{p},z) = \frac{1+\sigma}{E}\frac{e^{pz}}{p}\left[2\delta_{\alpha\beta} - (2\sigma - pz)\frac{p_\alpha p_\beta}{p^2}\right], \tag{A11a}$$

$$\chi^{st}_{z\beta}(\vec{p},z) = \frac{1+\sigma}{E}\frac{e^{pz}}{p}\, i\left[(1-2\sigma) - pz\right]\frac{p_\beta}{p}, \tag{A11b}$$

$$\chi^{st}_{\alpha z}(\vec{p},z) = \frac{1+\sigma}{E}\frac{e^{pz}}{p}\, i\left[-(1-2\sigma) - pz\right]\frac{p_\alpha}{p}, \tag{A11c}$$

$$\chi^{st}_{zz}(\vec{p},z) = \frac{1+\sigma}{E}\frac{e^{pz}}{p}\left[2(1-\sigma) - pz\right], \tag{A11d}$$

using the identity

$$\frac{1}{(2\pi)^2}\int d^2p\ e^{i\vec{p}\vec{r}}\frac{e^{pz}}{p} = \frac{1}{2\pi}\frac{1}{R} \quad (z \le 0) \tag{A12}$$

and its $z$ derivatives and integrations.

The quantity $\Phi_{ijk}(\vec{p},z)$ defined in (A8b) can be also obtained from Eqs. (A11) and the symmetry properties of the stifness tensor in an isotropic medium

$$c_{ijkl} = \frac{E}{2(1+\sigma)}\left[\delta_{ik}\delta_{jl} + \delta_{il}\delta_{jk} + \frac{2\sigma}{1-2\sigma}\delta_{ij}\delta_{kl}\right]. \tag{A13}$$

Here, we quote only the results needed,

$$\Phi^{st}_{z\alpha z}(\vec{p},z) = e^{pz}(-ip_\alpha)z, \tag{A14a}$$

$$\Phi^{st}_{zzz}(\vec{p},z) = e^{pz}(1-pz), \tag{A14b}$$

which allow the explicit multiplication

$$c_{az}(\vec{p}, z)\Phi_{zaz}(-\vec{p}, z) + c_{zz}(\vec{p}, z)\Phi_{zzz}(-\vec{p}, z) = \frac{1-s^{2}}{pE} f(p, s, z) \qquad \text{(A15)}$$

where

$$f(p, s, z) \equiv 2p \frac{e^{2pz}}{p}\left[1 - \frac{1}{1-s} pz + \frac{1}{1-s} p^{2} z^{2}\right], \qquad \text{(A16)}$$

or after the Fourier transformation with the help of Eq. (A12),

$$f(r, s, z) = \frac{1}{\sqrt{r^{2} + 4z^{2}}}\left[1 + \frac{1}{1-s}\frac{z^{2}}{r^{2} + 4z^{2}} + \frac{12}{1-s}\frac{z^{4}}{\left(r^{2} + 4z^{2}\right)^{2}}\right] . \qquad \text{(A17)}$$

Inserting (A15) into (A9), we finally obtain Eq. (4), namely

$$\operatorname{Im} c_{zz}^{w}(\vec{r}_1, \vec{r}_2) = \Delta f \frac{1-s^{2}}{pE}\left[f\left(\left|\vec{r}_1 - \vec{r}_2\right|, s, z_1\right) - f\left(\left|\vec{r}_1 - \vec{r}_2\right|, s, z_2\right)\right] . \qquad \text{(A18)}$$

## Appendix B

The expressions for the phase noise contain surface integrals of the form

$$I(\vec{r}_a,\vec{r}_b) = \iint dS' \iint dS'' \, e^{-2\left(\frac{(\vec{r}\,'-\vec{r}_a)^2}{w^2}\right)} e^{-2\left(\frac{(\vec{r}\,''-\vec{r}_b)^2}{w^2}\right)} f(\vec{r}\,'-\vec{r}\,'', z) \qquad \text{(B1)}$$

where the integrals over *dS'* and *dS"* are over the 2-dimensional plane of the mirror reflecting surface defined by vectors $\vec{r}\,'$ and $\vec{r}\,''$, and where *f* is defined as

$$f(\vec{r},z) \equiv \frac{1}{\sqrt{\vec{r}^{\,2} + 4z^2}} \left( 1 + \frac{1}{1-\sigma} \frac{z^2}{\vec{r}^{\,2} + 4z^2} + \frac{12}{1-\sigma} \frac{z^4}{\left(\vec{r}^{\,2} + 4z^2\right)^2} \right). \qquad \text{(B2)}$$

Due to the translational symmetry of the integration plane, the expression $I$ in Eq. (B1) is a function of the difference in the beam locations $\vec{r}_a - \vec{r}_b$. Indeed, by shifting the integration variables as $\vec{r}\,' \to \vec{r}\,' + \vec{r}_b$ and $\vec{r}\,'' \to \vec{r}\,'' + \vec{r}_b$, we can rewrite the expression for $I$ as

$$I(\vec{r}_a,\vec{r}_b) = I(\vec{r}_a - \vec{r}_b) = \iint dS' \iint dS'' e^{-2\left(\frac{(\vec{r}\,'-\vec{r}_a+\vec{r}_b)^2}{w^2}\right)} e^{-2\left(\frac{\vec{r}\,''^2}{w^2}\right)} f(\vec{r}\,'-\vec{r}\,'', z) \qquad \text{(B3)}$$

which not only proves the above assertion, but also shows that the integral is a double-convolution that can be evaluated easily in the Fourier space. In the lossy layer Green's function derivation in Appendix A, the function *f* was obtained from the Fourier representation

$$f(\vec{r},z) = \frac{1}{(2\pi)^2} \int d^2p \; e^{i\vec{p}\vec{r} - 2p|z|} \frac{4\pi}{2p} \left[ 1 + \frac{1}{1-\sigma} p|z| + \frac{1}{1-\sigma} p^2 z^2 \right] \qquad \text{(B4)}$$

Noting that the Fourier transform of a Gaussian is a Gaussian

$$\iint d^2r \, e^{-i\vec{p}\vec{r} - 2\vec{r}^{\,2}/w^2} = \iint d^2r \, e^{-2\left(\frac{\vec{r}}{w} + i\frac{w}{4}\vec{p}\right)^2 - \frac{w^2}{8}p^2} = \frac{\pi w^2}{2} e^{-\frac{w^2}{8}p^2}, \qquad \text{(B5)}$$

the expression for $I$ is given as an inverse Fourier integral

$$I(\vec{r}_a - \vec{r}_b) = \frac{1}{(2\pi)^2} \int d^2p \; e^{i\vec{p}(\vec{r}_a - \vec{r}_b)} \left( \frac{\pi w^2}{2} \right)^2 e^{-p^2 w^2/4} e^{-2p|z|} \frac{4\pi}{2p} \left[ 1 + \frac{1}{1-\sigma} p|z| + \frac{1}{1-\sigma} p^2 z^2 \right]. \qquad \text{(B6)}$$

The case of interest here is $\vec{r}_a = \vec{r}_b$, which simplifies the expression to

$$\begin{aligned} I &= \frac{1}{(2\pi)^2} \cdot 2\pi \cdot \left( \frac{\pi w^2}{2} \right)^2 \cdot 2\pi \int_0^\infty dp \; e^{-p^2 w^2/4 - 2p|z|} \left[ 1 + \frac{1}{1-\sigma} p|z| + \frac{1}{1-\sigma} p^2 z^2 \right] \\ &= \left( \frac{\pi w^2}{2} \right)^2 \cdot \frac{\sqrt{\pi}}{w} G(w,\sigma,z) \end{aligned} \qquad \text{(B7)}$$

where

$$\begin{aligned} G(w,\sigma,z) &\equiv \frac{w}{\sqrt{\pi}} \int_0^\infty dp \; e^{-p^2 w^2/4 - 2p|z|} \left[ 1 + \frac{1}{1-\sigma} p|z| + \frac{1}{1-\sigma} p^2 z^2 \right] \\ &= \frac{2}{\sqrt{\pi}} \int_0^\infty dx \; e^{-x^2 - 4(|z|/w)x} \left[ 1 + \frac{2}{1-\sigma} \frac{|z|}{w} x + \frac{4}{1-\sigma} \frac{z^2}{w^2} x^2 \right] \end{aligned} \qquad \text{(B8)}$$

The integral can be performed explicitly. Letting $u \equiv 2|z|/w$, we find that

$$G(w,s,z)=\frac{2}{\sqrt{p}}\int_0^{\infty} dx\ e^{-x^2-2ux}\left[1+\frac{1}{1-s}ux+\frac{1}{1-s}u^2x^2\right]. \qquad \text{(B9)}$$

With

$$I_1=\int_0^{\infty} dx\ e^{-x^2-2ux}=\frac{\sqrt{p}}{2}e^{u^2}\operatorname{erfc}(u) \qquad \text{(B10)}$$

$$I_2=\int_0^{\infty} dx\ e^{-x^2-2ux}x=\frac{1}{2}-\frac{\sqrt{p}}{2}ue^{u^2}\operatorname{erfc}(u) \qquad \text{(B11)}$$

$$I_3=\int_0^{\infty} dx\, e^{-x^2-2ux}x^2=-\frac{1}{2}u+\frac{\sqrt{p}}{2}\left(\frac{1}{2}+u^2\right)e^{u^2}\operatorname{erfc}(u) \qquad \text{(B12)}$$

we obtain

$$G(w,s,z)=\frac{2}{\sqrt{p}}\left[I_1+\frac{u}{1-s}I_2+\frac{u^2}{1-s}I_3\right]$$
$$=\frac{1}{\sqrt{p}}\frac{2}{1-s}\left[(z/w)-4(z/w)^3\right]+\left[1-\frac{2(z/w)^2}{1-s}+\frac{16(z/w)^4}{1-s}\right]e^{4(z/w)^2}\operatorname{erfc}(2z/w). \qquad \text{(B13)}$$

When $|z| << w$ if we keep terms up to linear first order in $|z|/w$ we find that

$$G(w,s,z)=1-\frac{2}{\sqrt{p}}\frac{1-2s}{1-s}\frac{|z|}{w}+O\left(\frac{z^2}{w^2}\right) \qquad \text{(B14)}$$

Figure 1: A plot of the function $G(w, s, z_1) - G(w, s, z_2)$ for $z_1 = 0$ as a function of $z_2 / w$ evaluated for the Poisson ratio of sapphire, taken to be 0.23.

Figure 2: A plot of $G(w, s, z_1) - G(w, s, z_2)$ for several values of $|z_2 - z_1|/w = \{0.1, 0.316, 1.0, 3.16, 10.0\}$ as a function of $z_1 / w$ evaluated for a Poisson ratio of 0.23.

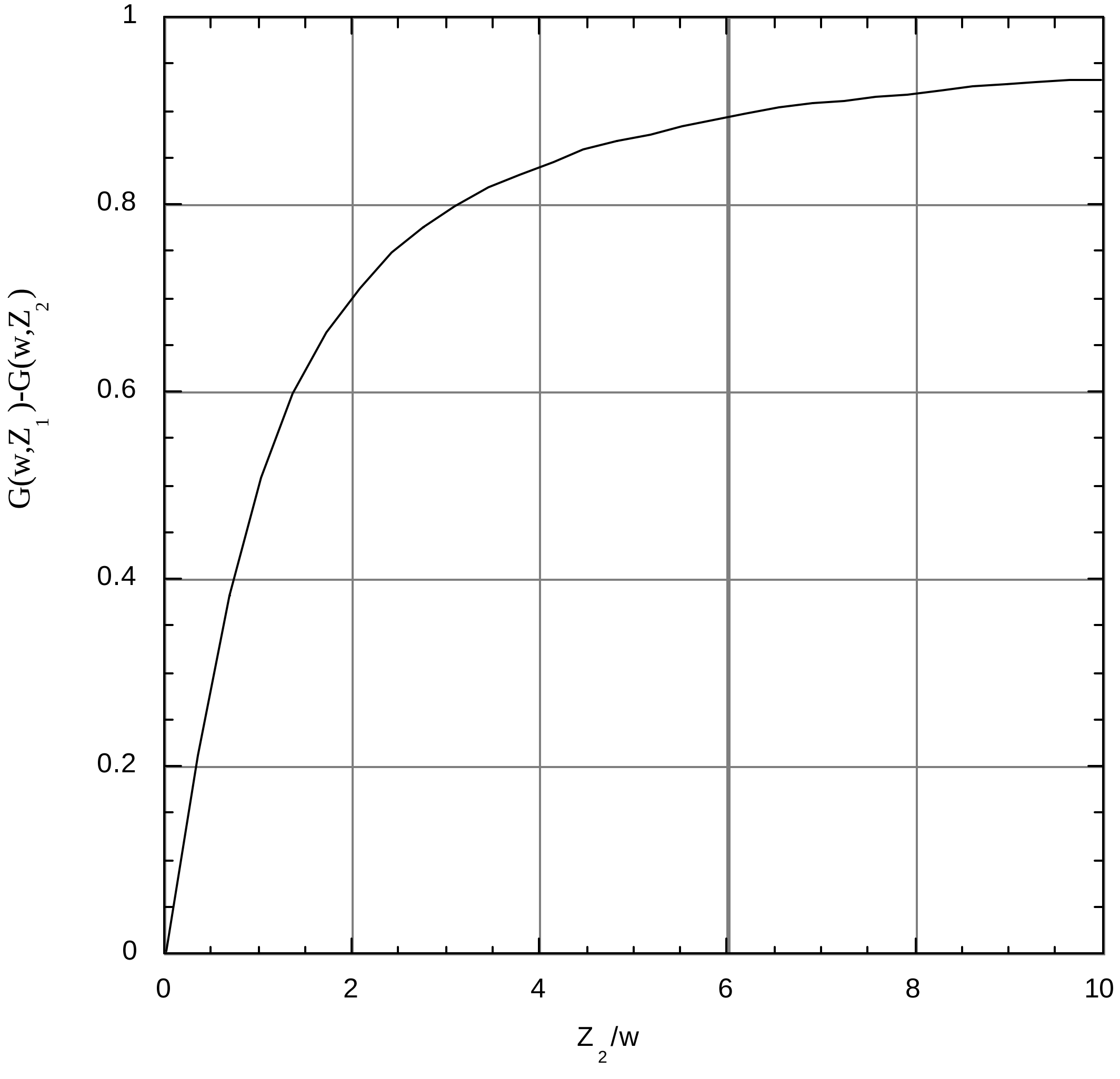

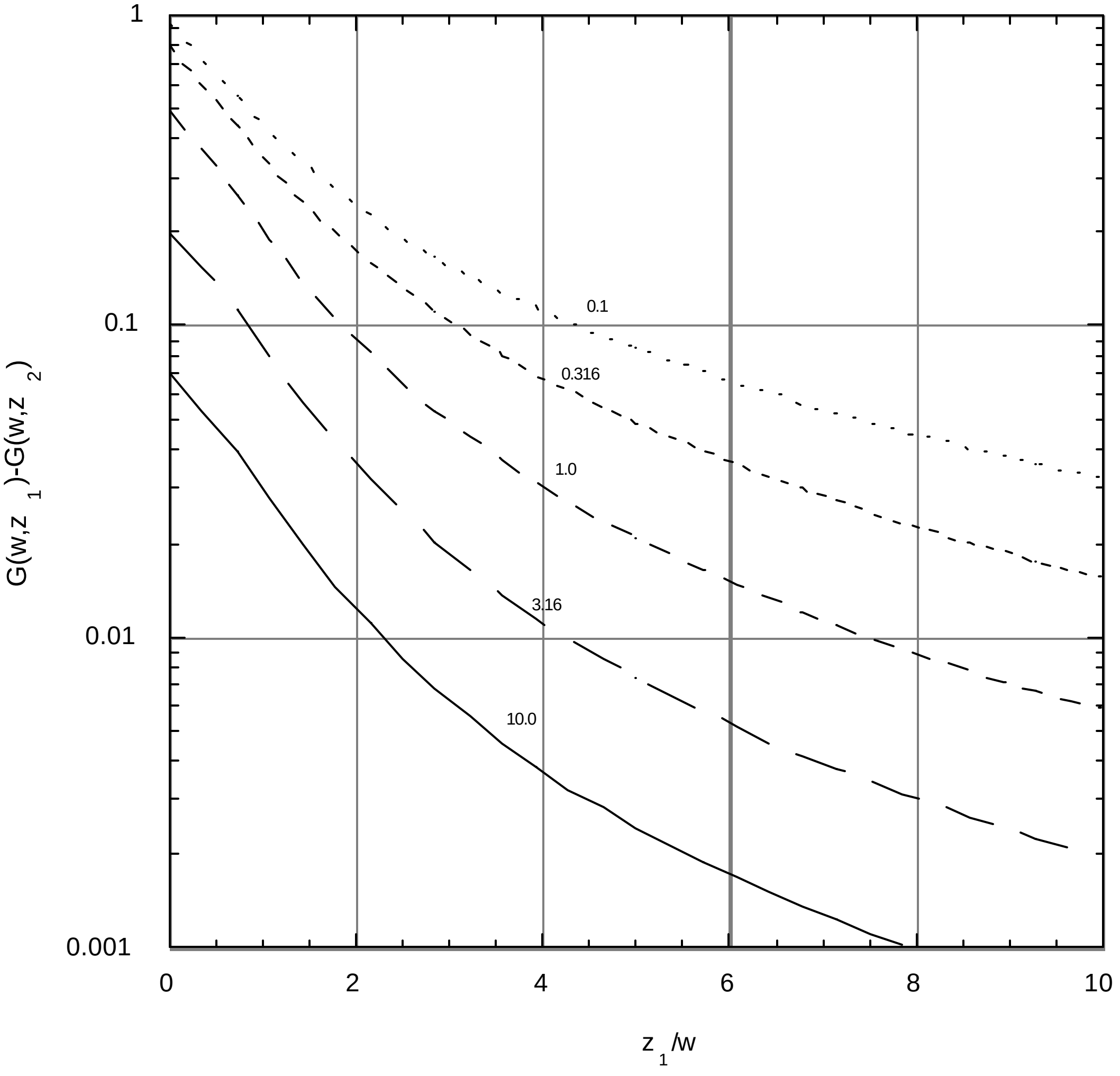